\documentclass[prl,aps,twocolumn,showpacs,superscriptaddress,floatfix]{revtex4}

\usepackage[latin1]{inputenc}
\usepackage{color}
\usepackage{graphicx,epsfig,epstopdf,amsmath}
\DeclareMathAlphabet{\mathpzc}{OT1}{pzc}{m}{it}
\begin{document}

%\title{Removing a magnetic framework dimension by pressure in CuF$_{2}$(D$_{2}$O)$_{2}$pyz: \\ from spin-wave to spinon excitations}

\title{Dimensional reduction by pressure in the magnetic framework material CuF$_{2}$(D$_{2}$O)$_{2}$pyz: from spin-wave to spinon excitations}

\author{M.~Skoulatos}
\affiliation{Heinz Maier-Leibnitz Zentrum (MLZ) and Physics Department E21, Technische Universit\"at M\"unchen, D-85748 Garching, Germany}
\affiliation{Laboratory for Neutron Scattering and Imaging, Paul Scherrer Institute, CH--5232 Villigen, Switzerland}

\author{M.~M\r{a}nsson}
\affiliation{Laboratory for Neutron Scattering and Imaging, Paul Scherrer Institute, CH--5232 Villigen, Switzerland}
\affiliation{Laboratory for Quantum Magnetism, \'{E}cole Polytechnique F\'{e}d\'{e}rale de Lausanne, Station 3, CH-1015 Lausanne, Switzerland}
\affiliation{Department of Materials and Nanophysics, KTH Royal Institute of Technology, SE-164 40 Kista, Sweden}

\author{C.~Fiolka}
\affiliation{Department of Chemistry and Biochemistry, University of Bern, Freiestrasse 3, 3012 Bern, Switzerland}

\author{K.~W.~Kr\"amer}
\affiliation{Department of Chemistry and Biochemistry, University of Bern, Freiestrasse 3, 3012 Bern, Switzerland}

\author{J.~Schefer}
\affiliation{Laboratory for Neutron Scattering and Imaging, Paul Scherrer Institute, CH--5232 Villigen, Switzerland}

\author{J.~S.~White}
\affiliation{Laboratory for Neutron Scattering and Imaging, Paul Scherrer Institute, CH--5232 Villigen, Switzerland}
\affiliation{Laboratory for Quantum Magnetism, \'{E}cole Polytechnique F\'{e}d\'{e}rale de Lausanne, Station 3, CH-1015 Lausanne, Switzerland}

\author{Ch.~R\"uegg}
\affiliation{Laboratory for Neutron Scattering and Imaging, Paul Scherrer Institute, CH--5232 Villigen, Switzerland}
\affiliation{Department of Quantum Matter Physics, University of Geneva, CH-1211 Geneva 4, Switzerland}

\date{\today}

\begin{abstract}

Metal organic magnets have enormous potential to host a variety of electronic and magnetic phases that originate from a strong interplay between the spin, orbital and lattice degrees of freedom. We control this interplay in the quantum magnet CuF$_2$(D$_2$O)$_2$pyz by using high pressure to drive the system through a structural and magnetic phase transition. Using neutron scattering, we show that the low pressure state, which hosts a two-dimensional square lattice with spin-wave excitations and a dominant exchange coupling of 0.89 meV, transforms at high pressure into a one-dimensional spin-chain hallmarked by a spinon continuum and a reduced exchange interaction of 0.43 meV. This direct microscopic observation of a magnetic dimensional crossover as a function of pressure opens up new possibilities for studying the evolution of fractionalised excitations in low dimensional quantum magnets and eventually pressure-controlled metal--insulator transitions.

\end{abstract}


\maketitle
Interacting spin systems with strong quantum fluctuations are an indispensable platform for exploring concepts in quantum many-body theory. For rigorous tests of theory it is often necessary to measure the system response due to external control parameters such as applied magnetic field and pressure ($P$). Just modest changes in these variables afford precise control over the strongly fluctuating degrees of freedom \cite{fr}, and provide a direct route towards exotic physics such as magnetic field- and $P$-driven quantum phase transitions \cite{ruegg,sebastian,merchant}. Another way to tune a system is to modify directly the magnetic superexchange, such as by changing the chemistry \cite{skoulatos,skoulatos2}, or by generating a sizeable in-situ modification of the crystal lattice. While the latter case proves difficult to achieve, it is much sought-after since it promises the direct control of magnetic dimensionality. Moreover, such an in-situ dimensionality change can be independent of the temperature, with such crossovers usually taking place as density of thermal fluctuations varies.

%Even though one can control the nature of magnetic superexchange by e.g. chemistry \cite{skoulatos,skoulatos2}

The different topologies of one-dimensional (1D) and two-dimensional (2D) quantum spin systems are well-known to define strongly contrasting ground states and dynamics. In 1D systems such as the half-integer spin chain, fermionic fractional excitations termed unbound spinons give a characteristic excitation continuum that disperses only along the chain direction \cite{ten}. In contrast, on the 2D magnetic square lattice, N\'{e}el-order and spin waves dispersing within the square lattice plane can be expected \cite{dalla}. Exploring an \textit{in-situ} transformation between 1D and 2D magnetic regimes is experimentally challenging due to lack of suitable materials, but it can be expected to provide both a deeper understanding of the adjacent states, and potentially the discovery of unusual phenomena in the vicinity of the crossover. For example, in recent work on 2D magnetic square lattice systems such as Cu(DCOO)$_2\cdot$4D$_2$O and La$_2$CuO$_4$, unusual features in the excitation spectra of each have been argued to reflect the existence of fractional excitations analogous to spinons in 1D systems \cite{dalla,headings}. Clarifying how fractional excitations evolve during a continuous 2D to 1D crossover can motivate new physical concepts, and at the same time provide links to related topics such as high-temperature superconductivity \cite{baskaran}. To progress the study of dimensional crossovers in quantum spin systems, simple materials with easily tunable structure-property relationships are required.\par

\begin{figure}[t!]
\centering
\includegraphics[width=\columnwidth]{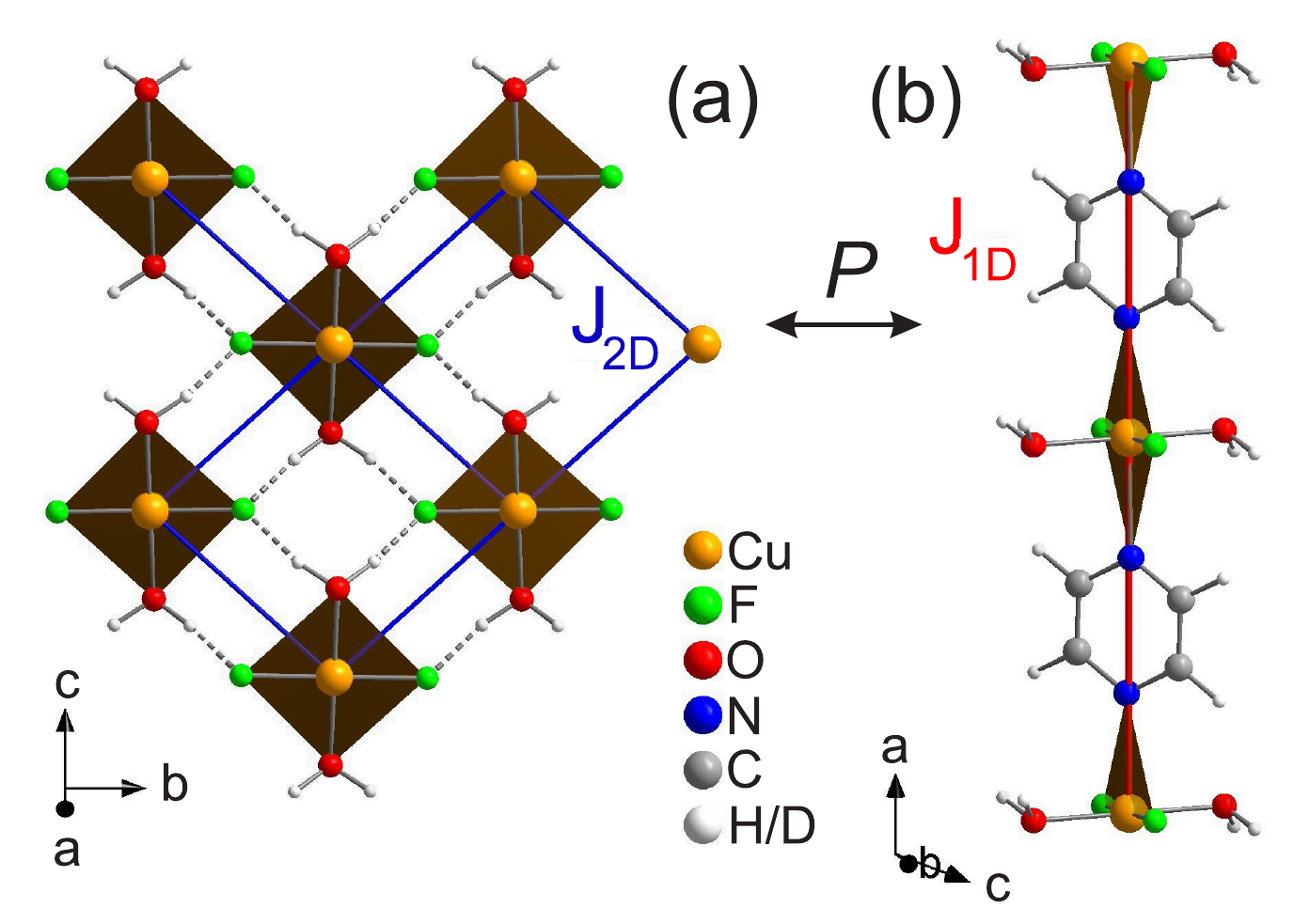}
\caption{CuF$_{2}$(D$_{2}$O)$_{2}$pyz crystal structure. (a) The $\alpha$-phase (ambient $P$) hosts 2D magnetic interactions on a slightly distorted square lattice. (b) The magnetic interactions switch to 1D in the $\beta$-phase ($P=$~6.1~kbar). The brown rhombi denote the half-occupied magnetic orbital of Cu$^{2+}$.}
\label{fig0}
\end{figure}

Here, neutron scattering has been used to explore a $P$-driven change of magnetic dimensionality in the metal organic compound CuF$_2$(D$_2$O)$_2$pyz, where pyz = D$_{4}$-pyrazine (Fig. 1). At $P=$~0 kbar, and in agreement with a previous report \cite{wang}, we find a simple antiferromagnetic (AFM) order below $T_N^{0kbar}=2.59$~K, where $S=1/2$ moments of 0.6 $\mu_B/$Cu$^{2+}$ lie in the $bc$ plane and are aligned along [0.7 0 1] in real-space. With well-defined spin-wave modes dispersing along $(0,Q_k,Q_l)$ and no dispersion along $Q_h$, the system displays typical spin-wave dynamics of a 2D magnetic square lattice with $J_{2D}^{0 kbar}$=0.89(2)~meV. Increasing $P$ to 3.6~kbar leads to a slight reduction in $T_N$ ($\Delta T=-0.13(1)$ K), and a small softening of the spin-wave modes ($J_{2D}^{3.6 kbar}$=0.78(3) meV). Further increasing pressure to 6.1~kbar leads to an entirely new phase where both the magnetic order and spin-wave modes of the 2D state are completely suppressed. Instead, a spinon continuum is observed along the $Q_h$ direction with $J_{1D}^{6.1 kbar}$=0.43(2) meV, which is the hallmark of a 1D spin chain. This remarkable $P$-driven magnetic dimensionality switch between 2D magnetic square lattice [Fig.~\ref{fig0}(a)] and 1D spin chain [Fig.~\ref{fig0}(b)] regimes is argued to be caused by a $P$-driven switch of a Jahn-Teller (JT) distortion characteristic to CuF$_{2}$(D$_{2}$O)$_{2}$pyz.\par

At ambient $P$, CuF$_{2}$(D$_{2}$O)$_{2}$pyz has a monoclinic structure ($P2_1/c$) with cell parameters of $a=7.6852$~\AA, $b=7.5543$~\AA, $c=6.8907$~\AA~and $\beta=111.25^{\circ}$ at room temperature (RT) [Fig.~\ref{fig0}(a)] \cite{halder1}. The magnetic Cu$^{2+}$ ions are coordinated by F$^{-}$ ions, O (from D$_{2}$O), and N atoms (from pyz), all in $trans$ positions of a slightly distorted octahedron with bond distances of Cu-F = 1.909 \AA, Cu-O = 1.969 \AA, and Cu-N = 2.397 \AA.
The Cu-F and Cu-O bonds are much shorter than the Cu-N bond, indicating that the half occupied magnetic $d_{x^2-y^2}$ orbital is localized in the $bc$ plane and the elongated Jahn-Teller (JT) axis parallel to the $a$-axis [Fig. 1(a)]. Deuterium bonds (or hydrogen bonds in the non-deuterated form \cite{note2}) F$\cdot\cdot\cdot$D-O provide an efficient superexchange pathway between the Cu$^{2+}$ ions and result in predominantly 2D magnetic interactions $J_{2D}$ on a slightly distorted square lattice \cite{halder1,godda}. Importantly, the different Cu-X (X=F,N,O) bond lengths make the JT axis an additional and controllable degree of freedom \cite{halcrow}.\par

Recently, CuF$_{2}$(D$_{2}$O)$_{2}$pyz was reported to show a series of $P$-driven, first-order structural phase transitions~\cite{halder1,presc,ghannadzadeh,lanza}. The ambient pressure phase, $\alpha$-CuF$_{2}$(D$_{2}$O)$_{2}$pyz switches to $\beta$-CuF$_{2}$(D$_{2}$O)$_{2}$pyz at $\sim9$~kbar, and $\gamma$-CuF$_{2}$(D$_{2}$O)$_{2}$pyz at $\sim32$~kbar. All of these phases exhibit the same crystal structure and maintain the monoclinic symmetry with spacegroup $P2_1/c$.
However, the elongated JT axis changes its orientation with increasing $P$ from the Cu-N to the Cu-O and Cu-F bond for the $\alpha$, $\beta$ and $\gamma$-phase, respectively.
$\beta$-CuF$_{2}$(D$_{2}$O)$_{2}$pyz has cell parameters of $a=6.8607$~\AA, $b=7.6549$~\AA, $c=7.0904$~\AA~and $\beta=114.289^{\circ}$ at RT and 12.39 kbar \cite{halder1}.
The corresponding bond distances are Cu-F = 1.955 \AA, Cu-O = 2.211 \AA, and Cu-N = 2.019 \AA. The elongated Cu-O bond locates the magnetic $d_{x^2-y^2}$ orbital in the $ab$ plane. The Cu-F$\cdot\cdot\cdot$D-O-Cu superexchange paths in the $bc$ plane are broken [Fig. 1(b)]. The switching of the JT axis results in a dramatic change of the magnetic properties between the $\alpha$ and $\beta$ phases.
While indeed bulk magnetic measurements indicate the $\alpha$-$\beta$ structural transition to separate very different magnetic regimes~\cite{halder1,ghannadzadeh}, direct microscopic characterisation of the associated magnetic states has been lacking up to now.\par

\begin{figure}[t!]
\centering
\includegraphics[width=\columnwidth]{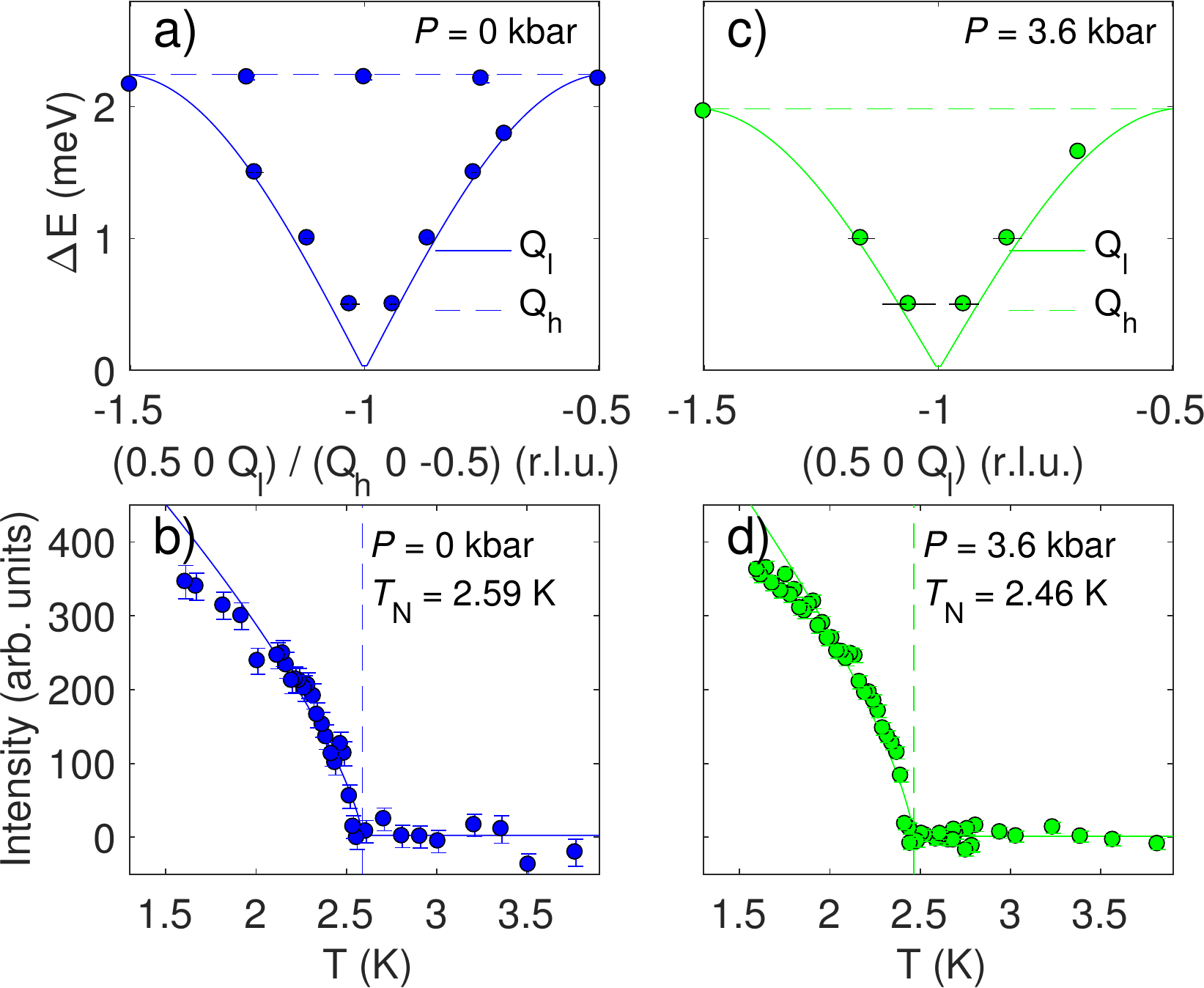}
\caption{Spin-wave excitations and magnetic order parameter of CuF$_{2}$(D$_{2}$O)$_{2}$pyz at $P=0$ kbar (a)-(b) and at $P=3.6$ kbar (c)-(d). (a) and (c) show typical 2D spin-wave dispersions obtained at 1.5~K, with the data fitted using linear spin wave theory. The solid (dashed) line are fits to the $Q_l$ ($Q_h$) dispersion. (b) and (d) show the $T$-dependence of the Bragg peak intensity at $\textbf{Q}=(1/2~0~0)$, with a power-law fit for the low $T$ region (see text).}
\label{fig1}
\end{figure}

Here we report high-$P$ neutron scattering studies of the microscopic magnetism in CuF$_{2}$(D$_{2}$O)$_{2}$pyz as pressure moves the system across the structural transition between the $\alpha$ and $\beta$ phases. Our experiments were done using two single-crystal pieces of deuterated CuF$_{2}$(D$_{2}$O)$_{2}$pyz (total mass 0.5~g) coaligned in the $(Q_h,0,Q_l)$ plane with a mosaic spread of 0.7$^\circ$. For high-$P$ measurements, the sample was loaded inside a piston-cylinder clamp cell. The pressure transmitting medium was a 1:1 mix of fluorinerts FC-75:FC-77. The applied pressure was monitored \textit{in-situ} by tracking the lattice compression of a NaCl crystal co-mounted with the sample, and using its reported equation of state \cite{skelton}. Inelastic neutron scattering (INS) data were collected using the TASP spectrometer at SINQ, PSI \cite{tasp}. Fixed outgoing wave vectors $k_f=1.3$ \AA$^{-1}$ and 1.5 \AA$^{-1}$ were used, along with a Be filter placed between the sample and analyzer in order to suppress higher-order contamination. All INS data were collected at $T=$~1.5~K. Further neutron diffraction data were collected on the single-crystal diffractometer TriCS at SINQ, using $\lambda=2.32$~\AA.

Fig.~\ref{fig1} shows a summary of the spin-wave dispersion and magnetic order in CuF$_{2}$(D$_{2}$O)$_{2}$pyz at $P=0$ kbar [Figs.~\ref{fig1}(a),(b)] and at $P = 3.6$ kbar [Figs.~\ref{fig1}(c),(d)]. At $P=$~0, AFM long-range-order onsets below $T_N^{0kbar}=2.59(1)$ K [Fig.~\ref{fig1}(b)].
The N\'eel temperature is suppressed at $P=$~3.6~kbar to $T_N^{3.6kbar}=2.46(1)$~K [Fig.~\ref{fig1}(d)]. In both cases, the magnetic peak intensity has been fitted with a power-law behavior for the temperature range $T>0.7*T_N$ and with a mean-field critical exponent fixed to the 3D Heisenberg case \cite{zinn}.
%Since at both pressures elastic magnetic intensity is observed at $\bf{Q}$=(1/2~0~0), this indicates that at $P=$~3.6~kbar the AFM moment directions of the magnetic square lattice remain aligned in the $bc$ plane diagonal as at $P=$~0.

\begin{figure}[t!]
\centering
\includegraphics[width=\columnwidth]{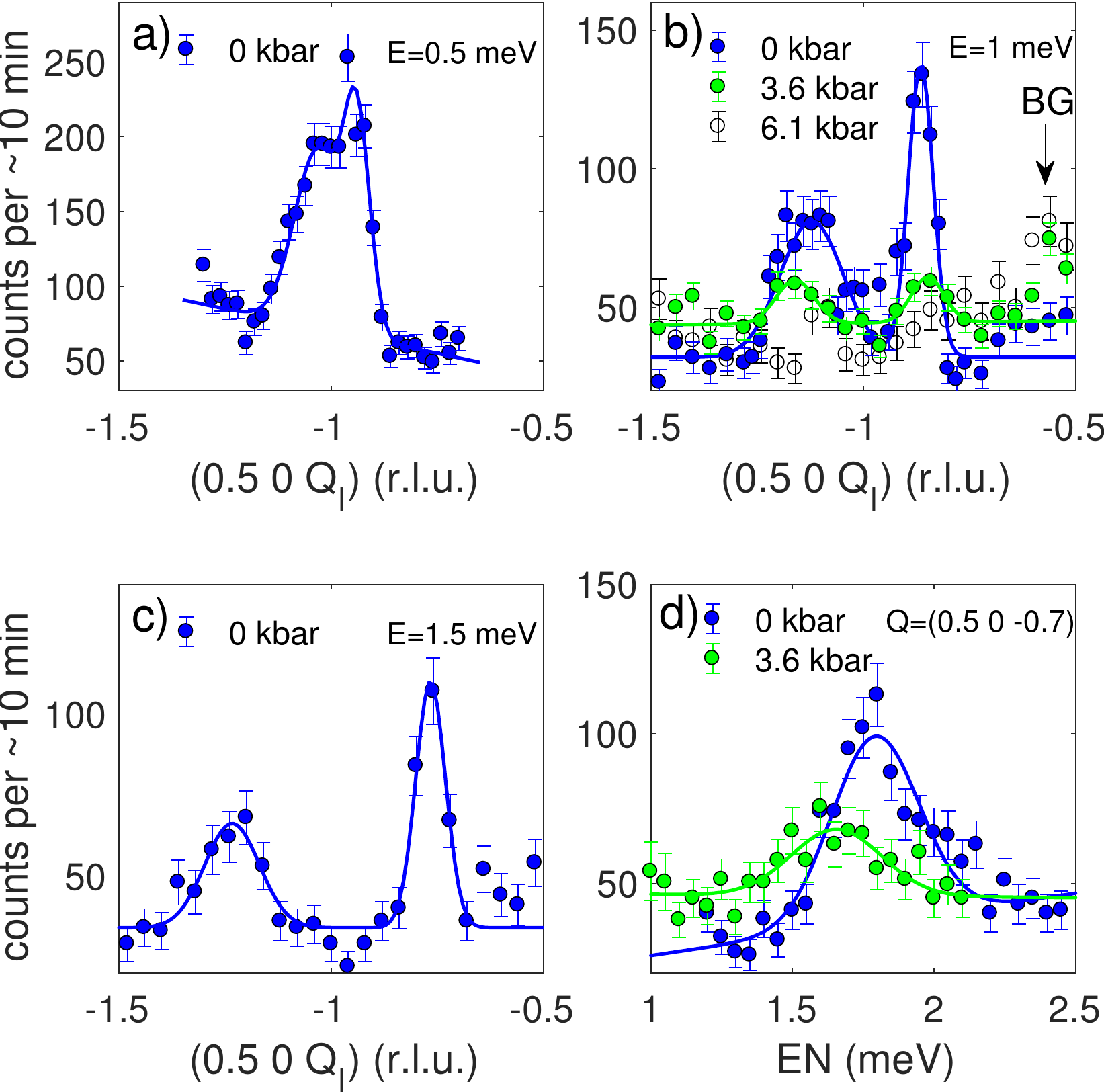}
\caption{INS data from CuF$_{2}$(D$_{2}$O)$_{2}$pyz at all measured pressures and 1.5~K. a-c) $Q$-scans at 0.5 meV, 1 meV and 1.5 meV energy transfers, respectively. d) shows energy-transfer scans at $Q$=(0.5 0 -0.7). The sharp spin-wave modes completely disappear at $P$=6.1 kbar (open points in b). Gaussian fits are used throughout (solid lines). The feature denoted as `BG' in panel (b) for $P>0$, is due to background from the pressure cell. It clearly disappears at $P=0$, where the cell was not in use. Note that the data of 0 and 3.6 kbar belong to the same phase (square lattice) and this serves as an independent criterion for this background.
%Note that the peak denoted as BG in Fig.~\ref{fig4}(a) is the same peak as in Fig.~\ref{fig2}(b) (same (Q,E) point due to the pressure cell). Correspondingly, it can also be seen in Fig.~\ref{fig3}(b) at an energy transfer of $\approx$ 1 meV.
}
\label{fig2}
\end{figure}

The reduction of $T_N$ by $\Delta T$ = -0.13(1)~K upon increasing pressure is accompanied by a softening of the spin waves, as shown in Figs.~\ref{fig1}(a) and (c). At both pressures, the magnetic excitations disperse away from the AFM Bragg point $\bf{Q}$=(1/2~0~-1), and along $Q_l$ in the magnetic square lattice plane. In contrast, an essentially flat dispersion is observed along the out-of-plane direction $Q_h$. Figure~\ref{fig2} shows in detail some of the energy-transfer- and $Q$-scans used to construct the dispersions shown in Fig.~\ref{fig1}. Figs.~\ref{fig2}(a)-(c) respectively show well-defined excitations in $Q_l$ scans at fixed energy transfers 0.5 meV, 1 meV and 1.5 meV. Fig.~\ref{fig2}(d) shows energy scans at fixed $\textbf{Q}$=(0.5~0~-0.7) for the two pressures. Here the shift down in energy of the peak at $P=$~3.6~kbar clearly shows a softening of the dispersion. A comparison between the two scans in Fig.~\ref{fig2}(d) also shows a reduced signal-to-noise ratio [see also Fig.~\ref{fig2}(b)], as a consequence of introducing the pressure cell.

Despite the slight softening of the excitation bandwidth at $P=$~3.6~kbar, the entire dispersion remains qualitatively similar. This strongly suggests $\alpha$-CuF$_{2}$(D$_{2}$O)$_{2}$pyz to host a regime of 2D magnetic interactions for all pressures in this structural phase. Using the SpinW library~\cite{sandor} we describe the dispersions using linear spin wave theory [see Figs.~\ref{fig1}(a) and (c)]. We find $\widetilde{J}_{2D}^{0kbar}=1.05(2)$ meV and $\widetilde{J}_{2D}^{3.6kbar}=0.92(3)$ meV for the 2D exchange coupling at each pressure, which become $J_{2D}^{0kbar}=0.89(2)$ meV and $J_{2D}^{3.6kbar}=0.78(3)$ meV after taking into account an appropriate quantum renormalisation factor $Z_c=1.18$ \cite{pelissetto,wang}. The result obtained at $P=$~0 kbar agrees well with previous work \cite{wang,ghannadzadeh,lanza}. Using our results for $J$ and T$_N$, we extract an empirical estimate \cite{yasuda} of the interlayer coupling $J_\perp \sim 3 \times 10^{-4}$~meV. This makes CuF$_{2}$(D$_{2}$O)$_{2}$pyz an excellent realization of a 2D square lattice AFM.\par

\begin{figure}[t!]
\centering
\includegraphics[width=\columnwidth]{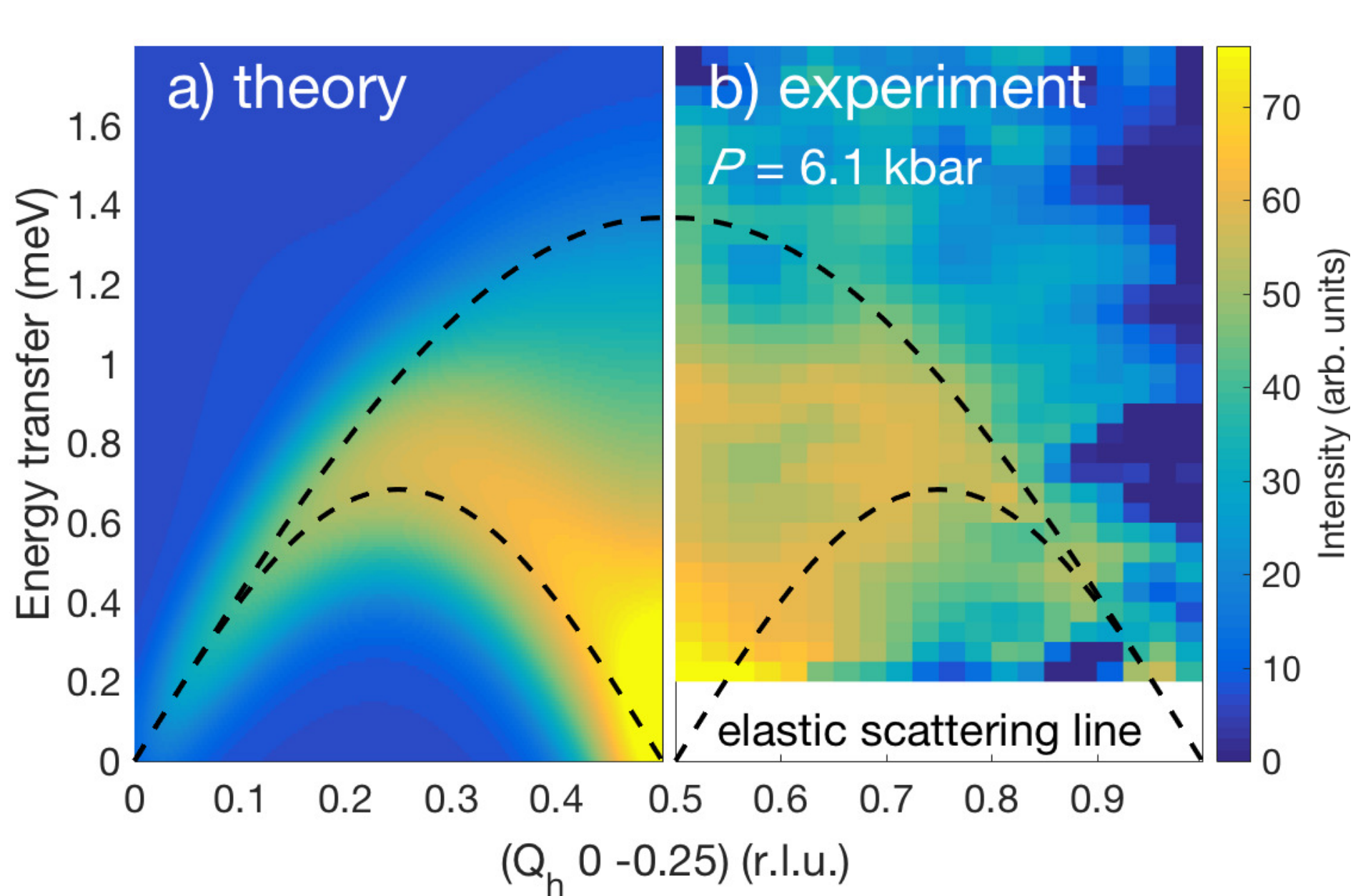}
\caption{Magnetic INS intensity maps of the spectrum from CuF$_{2}$(D$_{2}$O)$_{2}$pyz in the high-$P$ phase. (a) shows the theoretical spinon dynamic structure factor \cite{caux}. (b) shows the experimentally measured spectrum constructed from individual energy-transfer scans equally spaced in $Q$. The dashed lines correspond to the lower and upper spinon boundaries according to the `des Cloizeaux-Pearson' analytical result \cite{cloiz}.}
\label{fig3}
\end{figure}

\begin{figure}[t!]
\centering
\includegraphics[width=\columnwidth]{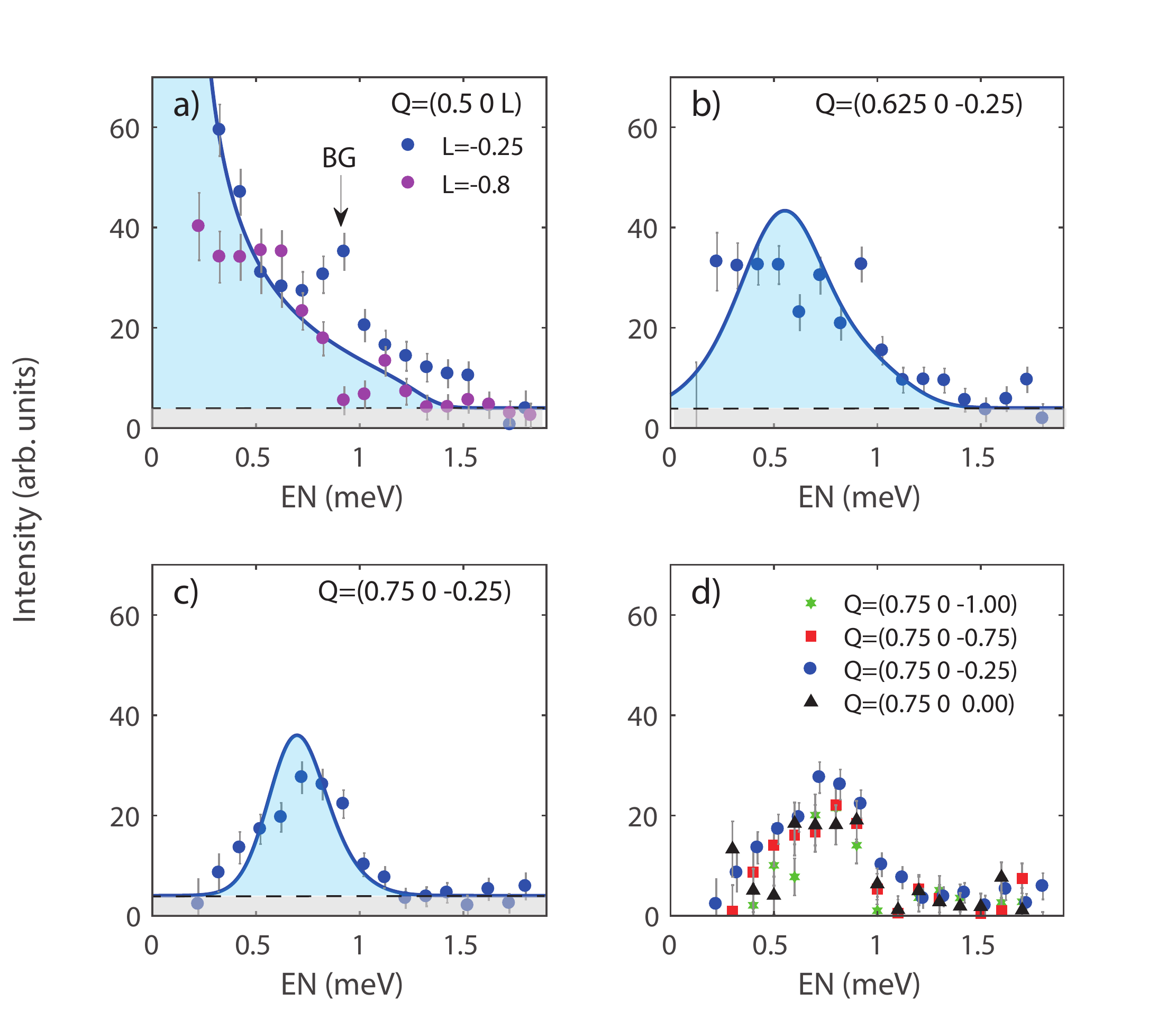}
\caption{Excitation spectra of CuF$_{2}$(D$_{2}$O)$_{2}$pyz in the 1D state. (a)-(c) show energy scans along the spin-chain direction $Q_h$. The solid lines are fits according to a theoretical spinon model ~\cite{caux}. (d) shows the spectrum to display no $Q_l$-dependence. The scans in all panels are obtained after subtracting an energy scan at $Q_h=0$ where no magnetic signal is expected [see Fig.~\ref{fig3}]. Despite this background subtraction, the signal denoted `BG' in (a) for $Q_L=-0.25$ is assigned to a parasitic, temperature-independent scattering process from the pressure cell as explained in the caption of Fig.~\ref{fig2}. Panel (a) includes an extra dataset at a different $Q_L$ value (=-0.8) with better conditions achieved, resulting to this background being absent.}
\label{fig4}
\end{figure}

At the highest $P$=6.1~kbar, our CuF$_{2}$(D$_{2}$O)$_{2}$pyz sample has transformed into the $\beta$ phase \cite{note3}. In this phase at 1.5~K, we found no evidence for either magnetic long-range-order or any sharp magnetic inelastic modes. This evidences the suppression of the low-$P$ 2D magnetic square lattice state, which has $Q_k$ and $Q_l$ as the main dispersion directions. Instead, fundamentally different behaviour is observed, since the INS response is manifested as a continuum of broad-scattering modes that disperse only along $Q_h$, and have no $Q_k$ or $Q_l$ dependence. The experimental data are summarised as an intensity color plot in Fig.~\ref{fig3}(b). The constituent energy scans used to construct Fig.~\ref{fig3}(b), are shown in Fig.~\ref{fig4}~\cite{footn}. From Figs.~\ref{fig4}(a)-(c) the dispersion along $Q_h$ is clearly seen, while data in Fig.~\ref{fig4}(d) show no significant dispersion along $Q_l$.

As can be seen in Fig.~\ref{fig3}, the spectrum observed from CuF$_{2}$(D$_{2}$O)$_{2}$pyz at $P=$~6.1~kbar [Fig.~\ref{fig3}(b)] changes dramatically and compares well with that expected for a spinon continuum dispersion of a 1D spin chain [Fig.~\ref{fig3}(a)] \cite{caux,mourigal}. The good agreement between experiment and theory is further emphasized when i) including the `des Cloizeaux-Pearson' analytical result \cite{cloiz} which defines the lower and upper spinon boundaries (dashed black lines in Fig.~\ref{fig3}), and ii) comparing the experimental data shown in Fig.~\ref{fig4}, and the spinon intensities modelled using the formalism of Ref.~\onlinecite{caux} (continuous lines in Fig.~\ref{fig4}(a)-(c)). From a simultaneous fit across all of our data, the spinon amplitude and $J_{1D}$ parameters were refined, yielding $J_{1D}^{6.1kbar}=0.43(2)$ meV.
%No $Q_l$ dependence of the spinon continuum is observed, see Fig. 5d).
This lies in agreement with the value $\sim 0.47$ meV used in Ref. \onlinecite{ghannadzadeh} to describe indirect bulk magnetization of the $\beta$-CuF$_{2}$(D$_{2}$O)$_{2}$pyz phase. Overall, our results unambiguously unveil the high-$P$ state of CuF$_{2}$(D$_{2}$O)$_{2}$pyz to bear the hallmark magnetic response of a 1D spin chain.

Experimentally we have observed a high-$P$ 1D magnetic state to be present in $\beta$-CuF$_{2}$(D$_{2}$O)$_{2}$pyz at $P=$~6.1~kbar. This pressure lies below $\sim$9~kbar reported in Refs.~\onlinecite{halder1,ghannadzadeh} for the $\alpha$-$\beta$ structural phase transition, yet within the range of 5-15~kbar reported in Ref.~\onlinecite{lanza}. A possible reason for any absolute pressure variations across different studies can be due to the use of different methods of pressure generation and related inhomogeneities. Another factor which may be important is our use of deuterated CuF$_{2}$(D$_{2}$O)$_{2}$pyz, since differences in properties could conceivably arise due to subtle changes between H and D bonding geometries \cite{godda,oneal}. While we cannot identify conclusively why the structural transition pressures vary between the different studies, we nevertheless observe directly a marked change in the system's magnetic and structural properties to occur over a pressure range that is broadly consistent with previous work \cite{halder1,ghannadzadeh,lanza}.

A consistent understanding of the pressure-driven change between the low-$P$ and high-$P$ magnetic regimes is achieved if the JT axis of the CuF$_2$O$_2$N$_2$ octahedra switches from the Cu-N bonds [Fig.~\ref{fig0}(a)] to the Cu-O bonds [Fig.~\ref{fig0}(b)]. This switch leads to a reorientation of the $d_{x^2-y^2}$ orbitals from the $bc$-plane at low-$P$ to the $ab$-plane at high-$P$ \cite{halder1}. Without microscopic magnetic measurements however, two competing scenarios for the high-$P$ magnetic state could nevertheless be suggested: i) the system in fact remains 2D, but with interactions in a different plane, or ii) 1D behaviour arises if a particular exchange path dominates over all others. Our direct neutron scattering measurements reveal the existence of a spinon continuum [Figs.~\ref{fig3}(b) and \ref{fig4}], confirming without doubt the 1D $S=1/2$ spin chain to be the correct magnetic model for the high-$P$ state.

Finally, we clarify the picture of the $P$-driven change in magnetic dimensionality in terms of the crucial exchange paths. In the 2D state, the Cu-F$\cdot\cdot\cdot$D-O-Cu bonds define the superexchange pathways. To be effective at mediating superexchange for a 2D square lattice topology, the near-orthogonal superexchange integrals traversing the $bc$ plane must be approximately equal. Once this $bc$-plane connectivity is suppressed, as happens in $\beta$-CuF$_{2}$(D$_{2}$O)$_{2}$pyz when the $d_{x^2-y^2}$ orbitals reorient to the $ab$-plane, the 1D spin chain state suggests that strongest superexchange pathways become those that are linear along the Cu-pyz-Cu network [Fig.~\ref{fig0}(b)]. In this case, the interaction $J_{1D}$ along the Cu-pyz-Cu bond is expected to be of the $\sigma$-type, and therefore weaker than for the 2D magnetic exchange in $\alpha$-CuF$_{2}$(D$_{2}$O)$_{2}$pyz. This picture is confirmed quantitatively from our data, since we find that $J_{1D}^{6.1~kbar}/J_{2D}^{0~kbar}\sim$~0.5.\par

In summary, our high pressure neutron scattering study of the metal organic magnet CuF$_{2}$(D$_{2}$O)$_{2}$pyz unveils a dramatic $P$-driven switch of the system's magnetic dimensionality. From neutron scattering measurements of the magnetic excitations, we report that the 2D magnetic square lattice state at low-$P$ transforms to a 1D spin chain at high-$P$. The microscopic mechanism behind this behaviour is argued to be a consequence of a $P$-driven switch of a uniquely sensitive Jahn-Teller axis in this soft material.
Here lies the essence of this work: this JT switch across another direction, has a direct consequence on the orbital occupancies and in turn to the exchange pathways, which are completely altering the magnetism of the system in a profound way.

More generally, our work provides a reference example of how dramatic effects on the magnetic states in a 3D coordination material can be generated via modest perturbations \cite{goddard1,lanza,oneal,presc}. Due to both the diverse range of metal organic materials with different ligands and networks that can be synthesized \cite{landee}, and their generally higher compressibility compared with inorganic compounds, high-$P$ becomes \emph{the} effective tool for seeking novel quantum states in these materials. In particular, the use of high-$P$ as a precision tuning parameter will be attractive for studying magnetic dimensionality changes across a \emph{continuous} $P$-driven structure transition, as opposed to the first-order structural transition which takes place in CuF$_{2}$(D$_{2}$O)$_{2}$pyz. In the former case, the coupling between strong quantum magnetic and enhanced structural fluctuations in the crossover region may lead to exotic quantum magnetic and magnetoelastic states~\cite{busch,brinz}.
An ultimate goal may also be to control a metal-insulator transition in such systems.

\textit{Acknowledgements:} We are grateful to J.S.~Caux for communicating his theoretical results on spinon intensities, D.~Sheptyakov for help during the experiment, and S.~Ward for fruitful discussion. MS acknowledges funding from the European Community's Seventh Framework Programme (FP7/2007-2013) under Grant Agreement No. 290605 (PSI- FELLOW/COFUND) as well as TRR80 of DFG. The financial support by the Swiss National Science Foundation is gratefully acknowledged (Grant No. 200020-150257). This work is based on experiments performed at the Swiss spallation neutron source (SINQ), Paul Scherrer Institute (PSI), Villigen, Switzerland.


\begin{thebibliography}{00}
\bibitem{note2} In our experiments we study the deuterated form of CuF$_{2}$(D$_{2}$O)$_{2}$pyz. The deuteration of hydrogen-containing compounds to be studied by neutron scattering is standardly done since it significantly reduces the incoherent neutron scattering. Consequently, for deuterated samples the background scattering detected is suppressed as compared with hydrogenated samples.
\bibitem{note3} From our neutron experiments at TASP and at $T=$~1.5~K, the crystal parameters accessible in the horizontal $[H0L]$ plane were determined at each $P$ (to accuracy 1 in the last digit). For $P=$~0: a~=~7.61~\AA, c~=~6.83~\AA, $\beta=$~112.1$^{\circ}$, for $P=$~3.6~kbar: a~=~7.60~\AA, c~=~6.79~\AA, $\beta=$~113.3$^{\circ}$, and for $P=$~6.1~kbar: a~=~6.87~\AA, c~=~7.09~\AA, $\beta=$~113.4$^{\circ}$. The sharp change in $a$ and $c$ lattice parameters on going from $P=$~3.6~kbar to $P=$~6.1~kbar are consistent with the occurrence of the $\alpha$-$\beta$ structure transition, reported to take place at $\sim9$~kbar at RT~\cite{halder1,lanza}.
\bibitem{footn} As a consequence of using the $P$-cell, various reciprocal lattice $Q_k,~Q_l$ points were tested in terms of optimising the signal-to-noise for the measurements. The $Q_h$ dependence was optimally conducted at $Q_k=0$ and $Q_l=-0.25$.
\end{thebibliography}
\end{document}